\documentclass[showpacs,amsmath,amssymb,amsfonts,pre,floatfix,twocolumn,superscriptaddress]{revtex4} %superscriptaddress
\usepackage{bm}
\usepackage[latin1]{inputenc}
\usepackage[english]{babel}
\usepackage[T1]{fontenc}
\usepackage[dvips]{graphicx}
\usepackage{color}
\usepackage{subfigure}
\date{\today}

\begin{document}

\title{Few-boson tunneling in a double well with spatially modulated interaction}

\author{Budhaditya Chatterjee}
\email{bchatter@PHYSnet.uni-hamburg.de}
\affiliation{Physikalisches Institut, Universit\"at Heidelberg, Philosophenweg 12, 69120 Heidelberg, Germany}

\author{Ioannis Brouzos}
\email{ibrouzos@PHYSnet.uni-hamburg.de}
\affiliation{Zentrum f\"ur Optische Quantentechnologien, Luruper Chaussee 149, 22761 Hamburg, Germany}

\author {Sascha Z\"ollner}
\email{zoellner@nbi.dk}
\affiliation{Niels Bohr International Academy, Niels Bohr Institute, Blegdamsvej 17, 2100 K\o benhavn, Denmark}

\author{Peter Schmelcher}
\email{Peter.Schmelcher@PHYSnet.uni-hamburg.de}
\affiliation{Zentrum f\"ur Optische Quantentechnologien, Luruper Chaussee 149, 22761 Hamburg, Germany}

\begin{abstract}
We study few-boson tunneling in a one-dimensional double well with a spatially modulated interaction. The dynamics changes from Rabi oscillations in the non-interacting case to a highly suppressed tunneling for intermediate coupling strengths followed by a reappearance near the fermionization limit. With extreme interaction inhomogeneity in the regime of strong correlations we observe tunneling between the higher bands. The dynamics is explained on the basis of the few-body spectrum and stationary eigenstates. For higher number of particles, $N\geq 3$ it is shown that the inhomogeneity of the interaction can be tuned to generate tunneling resonances. Finally, a tilted double-well and its interplay with the interaction asymmetry is discussed.
\end{abstract}

 \maketitle

\section{Introduction}

Ever since the experimental realization of Bose-Einstein condensation, ultracold atoms have been used to study an enormous diversity of quantum effects with an unprecedented degree of control \cite{pitaevskii,pethick}. Advancement in tools like optical lattices \cite{bloch07}  offer us the  chance to explore  e.g. second-order tunneling \cite{foelling07}, quantum phase transitions \cite{greiner02} and non-equilibrium quantum dynamics of driven systems \cite{kierig08}.

The double well especially  serves as a prototype system  to study interference and tunneling in great detail. For instance the tunneling dynamics of a Bose-Einstein condensate has been observed to undergo Josephson oscillations \cite{albiez05,milburn97,smerzi95} in which the population simply tunnels back and forth between the two wells. However when the interaction is raised beyond a critical value the atoms remain trapped in one well, a non-linear phenomenon known as  self trapping \cite{albiez05,smerzi95,anker05}.

Of special interest are systems in lower dimensions which often display unique features. Quasi-one-dimensional (1D) Bose gases have been prepared experimentally by freezing the transverse degrees of freedom. There it is possible to tune the interaction strength between the atoms by either using confinement induced resonances \cite{Olshanii1998a} or magnetic Feshbach resonances \cite{koehler06}. Thus one can study the crossover from a weakly interacting to a strongly correlated regime.
  A particularly interesting case  is the Tonks-Girardeau gas appearing in 1D in the limit of infinitely repulsive short-ranged interaction which has been recently observed experimentally \cite{kinoshita04,paredes04}. This gas of impenetrable boson is isomorphic with that of  free fermions via the Bose-Fermi mapping \cite{girardeau60} and all the local properties are identical to the free fermion system. The gas still retains the bosonic permutation symmetry and so the non-local quantities differ from the fermionic case.

Theoretically the quantum dynamics in the weakly interacting case has been studied using the Bose-Hubbard model assuming the validity of a lowest band approximation \cite{salgueiro06,dounasfrazer07a,dounasfrazer07b,wang08}. These studies illuminate relevant tunneling mechanisms and resonances. However, to capture the rich physics present in the stronger interaction regime we need to go beyond the Bose-Hubbard limit. Moreover numerically exact calculations of the quantum dynamics for few bosons through a one-dimensional potential barrier \cite{alexej08} or a bosonic Josephson junction  \cite{sakmann09} reveal deviations from the results obtained with mean-field calculations   as well as establish a difference between the dynamics in attractive and repulsive bosonic systems \cite{sakmann10}. The crossover from the uncorrelated to fermionization regime has been investigated for few bosons \cite{zoellner07a,zoellner08} and reveals a transition from Rabi-oscillations to fragmented pair tunneling via a highly delayed tunneling process  analogous to the self-trapping for condensates. The quantum dynamics of an asymmetric double-wells while keeping a constant interaction strength has been explored in refs. \cite{dounasfrazer07a,dounasfrazer07b,zoellner07a,zoellner08}.

In this investigation we go one step further and envision a new approach to asymmetry by  introducing an inhomogeneous, i.e., spatially varying interaction strength. This can be achieved experimentally  by employing magnetic field gradients in the vicinity of Feshbach resonances or by combining magnetic traps with optically induced Feshbach resonances \cite{koehler06,bauer09}. We will demonstrate how spatially varying interaction strengths enrich the tunneling dynamics in the most fundamental case of a double-well. This has to be seen as a potential ingredient for more complex problems such as quantum transport in optical lattices. Specifically we study the crossover from the non-interacting to the fermionization limit for a fixed inhomogeneity ratio of interaction and the effect of varying inhomogeneity ratio. An interplay of suppression and resumption of tunneling is observed. For three or more particles, the interaction asymmetry can be tuned to generate various many-particle tunneling resonances. Lastly we examine a tilted double-well and  investigate the interplay between the tilt and inhomogeneity to generate tunneling resonances.

The paper is organized as follows. In Section \ref{sec:setup} we discuss our model and setup. In Sec. \ref{sec:mctdh} we briefly describe our computational method. Subsequently we present and discuss the results for tunneling in a symmetric double well for two atoms  first (Sec. \ref{sec:2p_dynamics}) followed by more atom systems (Sec. \ref{sec: many particle}). In Sec. \ref{sec:asymmetric} we discuss the case of an asymmetric double well.

\section{Setup and Interactions\label{sec:setup}}

 Our Hamiltonian (for $N$ particles) is given by (see \cite{zoellner06a} for details) \\

\begin{equation}H = \sum_{i=1}^N [\frac{1}{2} {p_i}^2 + U(x_i)] + g\sum_{i<j} \delta(x_i - x_j) \end{equation}
The double well
$U(x) = \frac{1}{2} x^2 + h\delta_{\omega} (x)$
is modeled as a harmonic potential with a central barrier shaped as a Gaussian  $\delta_{\omega}
(x) = \frac{e^{-x^2/2\omega^2}}{\sqrt{2\pi}\omega}$ (of width $\omega = 0.5 $ and height $h = 8$, where dimensionless harmonic-oscillator units are employed  throughout).

For ultracold atoms only the s-wave scattering is relevant and the effective interaction in 1D  can be written as a contact potential \cite{Olshanii1998a} which we sample here by a very narrow Gaussian.
 We focus on repulsive interaction only.

The inhomogeneity of the interaction is modeled as \cite{zoellner06a} \\
\begin{center}
$g(R)= g_0[1 + \alpha \tanh (\frac{R}{L})]$, \\
\end{center}
where $2R=x_i + x_j$ and $L$ is the modulation length which we fix at $L = 1$. \\
 For $R\gg L$, $g$ takes the asymptotic  values \\
\begin{center}
$g_{\pm} = g_0 (1\pm \alpha)$.
\end{center}
Thus the parameter  $\alpha$ regulates the relative difference in interaction strength between the left and the right well,
\begin{center}
$\Delta g \equiv \vert g_+ - g_-\vert = 2g_0 \alpha$, \\
\end{center}
and the corresponding ratio is given by
\begin{center}

$\frac{g_+}{g_-} =\frac{1+\alpha}{1-\alpha} $. \\

\end{center}

\section{Computational Method\label{sec:mctdh}}
Our goal is to study the bosonic quantum dynamics for weak to strong interactions in a numerically exact fashion. This is computationally challenging and can be achieved only for few atom system.
Our approach is the Multi-Configuration Time Dependent Hartree (MCTDH) method \cite{meyer90,beck00} being a wave packet dynamical tool known for its outstanding efficiency in high dimensional applications.

The principle idea is to solve the time-dependent Schr\"odinger equation \\
\begin{center}
$i\dot{\Psi}(t) = H\Psi(t)$ \\
\end{center}
as an initial value problem by expanding
the solution in terms of Hartree products $\Phi_J \equiv {\varphi_j}_1 \otimes \ldots \otimes {\varphi_j}_N$ :

\begin{equation}\Psi(t) = \sum_J A_J(t)\Phi_J(t).\label{eq:mctdh}\end{equation}

The unknown single particle functions $\varphi_j(j=1,...,n$, where $n$ refers to the total number of single particle functions used in the calculation) are in turn  represented in a fixed primitive basis implemented on a grid. The correct bosonic permutation symmetry is obtained by symmetrization of the expansion coefficient $A_J$.
Note that in the above expansion, not only are the coefficients $A_J$ time dependent but  also the single particle functions $\varphi_j$.
Using the  Dirac-Frenkel variational principle, one can  derive the equations of motion for both $A_J$ and $\Phi_J$.
Integrating these differential equations of motion  gives us  the time evolution of the system via (\ref{eq:mctdh}).
This has the advantage that the basis $\Phi_J(t)$ is variationally optimal at each time $t$. Thus it can be kept relatively small, rendering the procedure more efficient.

Although MCTDH is designed primarily for time dependent problems, it is also possible to compute  stationary states. For this purpose the \textit{relaxation} method is used \cite{kos86:223}. The key idea is to propagate a wave function $\Psi_0$ by the non-unitary operator $e^{-H\tau}$. As $\tau \rightarrow \infty$, this exponentially damps out any contribution but that stemming from  the true ground state like $e^{-(E_m - E_0)\tau}$. In practice one relies upon a more sophisticated scheme called the \textit{improved relaxation} \cite{mey03:251,meyer06} which is much more robust especially for excited states.  Here $\langle\Psi\vert H \vert \Psi \rangle$ is minimized with respect to both the coefficients $A_J$ and the orbitals $\varphi_j$. The effective eigenvalue problems thus obtained are then solved iteratively by first solving $A_J$ with fixed orbital $\varphi_j$ and then optimizing $\varphi_j$ by propagating them in imaginary time over a short period. This cycle is then repeated.

\section{Tunneling Dynamics for Two Boson System\label{sec:2p_dynamics}}

We first focus on the tunneling dynamics in a symmetric double-well with two bosons initially ($t=0$) prepared in the left well. This is achieved by adding a tilt or a linear potential $dx$ to the Hamiltonian hence making the left well energetically favorable. Instantaneously  the ground-state is obtained by applying the relaxation method (imaginary time propagation). For reasonably large $d$, this results in achieving a complete population imbalance between the wells. With this state as the initial state, the tilt is instantaneously ramped down ($d=0$) at $t=0$ to study the dynamics in a symmetric double-well. Our aim is to study the impact of the correlations between the bosons on the tunneling dynamics both with respect to the interaction strength as well as the spatial inhomogeneity. We start by fixing the inhomogeneity to $\alpha = 0.2$ (with the left well having lower interaction than the right) and analyze how the dynamics varies with changing interaction strength $g_0$.

\subsection{Dynamics from the uncorrelated to the fermionization limit.\label{sub:2p-cor_dynamics}}

\begin{figure}
\includegraphics[width=0.95\columnwidth,keepaspectratio]{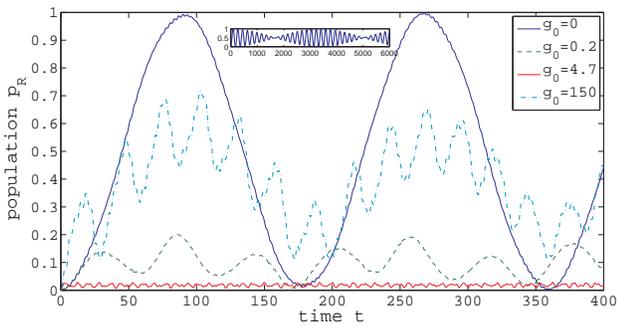}
\caption{(color online)  Population of the right-hand
well over time, $p_{\mathrm{R}}(t)$, for different interaction strengths at $\alpha = 0.2$ for two bosons. \textit{Inset}: Long time behavior for very low interaction strength $g_0=0.005$. Barrier height $h=8$ and width $\omega = 0.5 $  has been used for all calculations. (all quantities are in dimensionless harmonic oscillator units throughout).\label{cap:2p_g_var}}
\end{figure}

In the absence of any interaction $g_0=0$, the bosons undergo Rabi oscillations between the two wells. This is characterized by complete tunneling of both bosons between the two wells with a single frequency and  can be quantified  by the time variation of the population of the atoms in the right well
\begin{eqnarray*}P_R (t) = \langle \Theta (x) {\rangle}_{\Psi(t)} = {\int_0}^\infty \rho(x;t)dx \end{eqnarray*}  where $\rho$ is the one-body density.
Figure \ref{cap:2p_g_var} shows that $P_R$ oscillates sinusoidally between $0$ and $1$.
If we introduce a very small interaction $g_0 = 0.005$ (inset) the Rabi oscillations  give way to a beat pattern due to the existence of two very close frequencies.
Increasing the interaction strength further ($g_0=0.2$), we observe a suppression of tunneling with the maximum population in the right well ${P_R}^{max}\approx 0.2$. This is a manifestation of the inhomogeneous interaction which drives the tunneling off-resonance and should be carefully distinguished from the delayed pair-tunneling and self-trapping for the same $g_0$-value in the case of homogeneous interactions (see below). The dynamics consists of a slow tunneling envelope with suppressed amplitude, which is modulated by a faster oscillation. For higher values of interaction strength $(g_0 = 4.7)$, the tunneling is completely suppressed. What remains is a fast oscillation with a tiny amplitude.

However, contrary to the naive intuition a  reappearance of tunneling occurs for larger values of the coupling strength. We observe a partial restoration of tunneling with ${P_R}^{max} = 0.7$ for the value $g_0=150$, which is close to the so called fermionization limit. The dynamics is characterized by two frequencies - one very close to the Rabi frequency modulated by a faster oscillation. Ideally at the fermionization limit $g_0 \rightarrow \infty$, the system of hardcore bosons maps to a system of free fermions \cite{girardeau60} and all the local properties are identical. Hence in this limit we would have complete two-mode single particle tunneling analogous to tunneling of two free fermions.

Before we move on to analyze in detail the above observations, let us comment briefly on the differences between the behavior observed in our setup and the case of a symmetric double-well with homogeneous interaction. Clearly both for the non- and infinitely interacting limits the inhomogeneity doesn't play a role. For homogeneous interactions and a symmetric trap  tunneling is always resonant and complete. However,  different strength of interaction yield different  dynamics like a transition from pair-tunneling for low interaction strength to a self-trapping mechanism for larger interaction strength which is characterized by extremely long tunneling times \cite{salgueiro06,tonel05,creffield07,zoellner07a,zoellner08}. In our case though we observe an actual suppression of the tunneling amplitude and not so much a delayed process.
In case of an asymmetric-well with homogeneous interaction, the effects in the low interaction regime are equivalent to our set-up: The tilt has the same effect as an interaction asymmetry, namely it destroys resonant behavior thereby leading to a suppression of tunneling \cite{dounasfrazer07a,dounasfrazer07b}. Nevertheless, our case is fundamentally different and this is evident in the strong interaction regime. Specifically the reemergence of tunneling we observe does not occur in the tilted double-well system.

\subsection{Analysis\label{sub:analysis}}

\begin{figure}

\includegraphics[width=0.85\columnwidth,height=4.5cm]{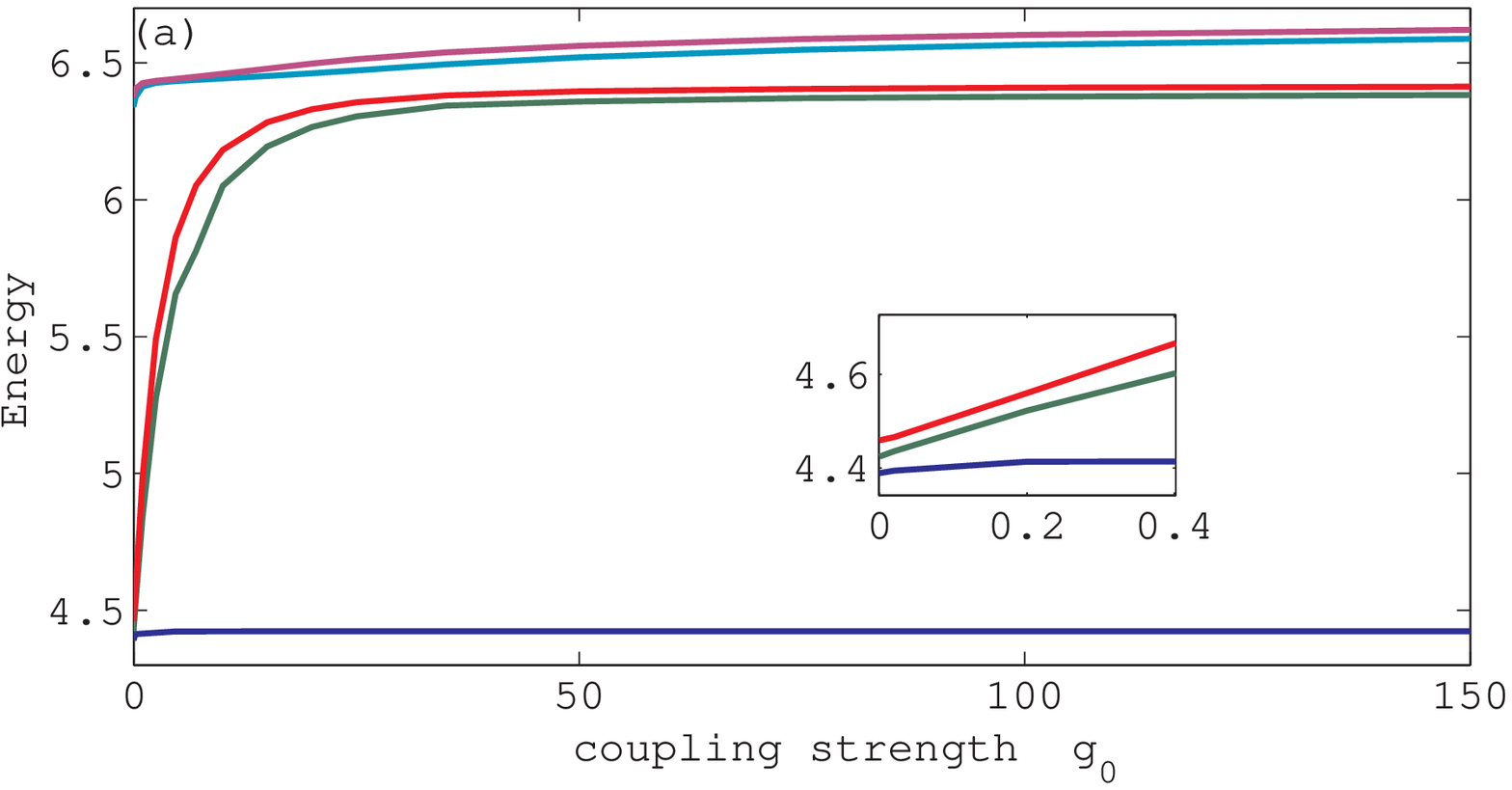}
\includegraphics[width=0.45\columnwidth,keepaspectratio]{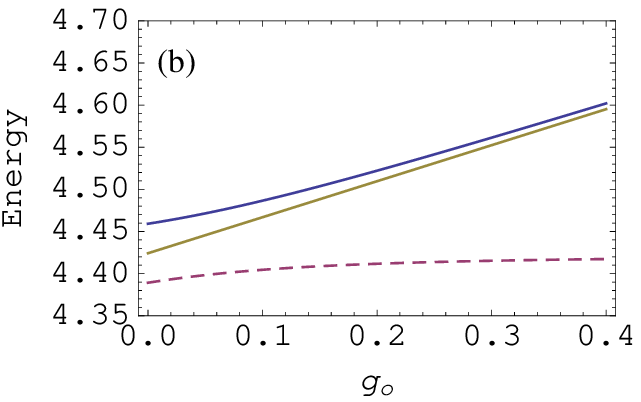}
\includegraphics[width=0.45\columnwidth,keepaspectratio]{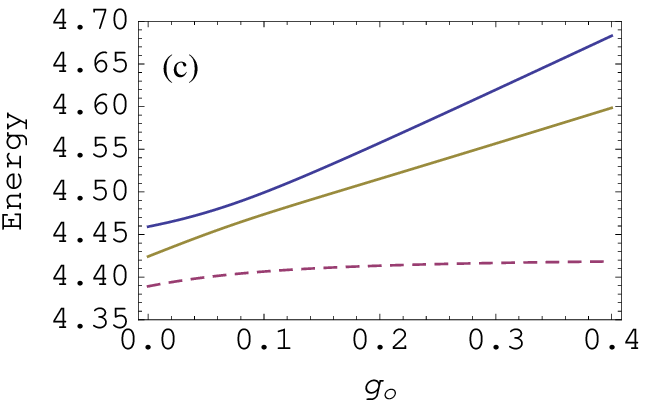}

\caption{(color online):(a) Two particle energy spectrum as a function of the interaction strength $g_0$ for $\alpha = 0.2$. \textit{Inset}: Lowest energy levels for low interaction strength.
\textit{Bottom}. Few body energy spectrum with $g_0$ for (b) $\alpha$ = 0 and (c)  $\alpha$ = 0.2.
\label{cap:spectrum}}

\end{figure}

The understanding of the above-described dynamics lies in the variation of the few body spectrum as $g_0$ is changed from zero to the fermionization limit (Fig.\ref{cap:spectrum}(a)).
Considering the  wavefunction $\Psi(t)= \sum_m e^{-iE_mt}c_m \Psi_m$ with energy $E_m$ corresponding to the stationary state $\Psi_m$, the population imbalance $\delta(t) \equiv \langle\Theta(x) - \Theta(-x)\rangle_{\Psi(t)} $ can be computed to be \\
\begin{equation}\delta(t) = 4\sum_{m<n} {W_m}_n \cos ({\omega_m}_n t) + 2\sum_m {W_m}_m -1 \label{eq:contribution},\\
\end{equation}
where
${W_m}_n = \langle \Psi_m\vert\Theta(x)\vert\Psi_n\rangle c_m c_n$ and ${\omega_m}_n = E_m -E_n$. \\

The energy spectrum of both the non-interacting and the fermionization limit can be understood from the single particle energy spectrum of the double well, which is in the form of bands each pertaining to a pair of symmetric and antisymmetric orbitals.

In the uncorrelated limit ($g_0 \rightarrow 0$) the low-lying energies of the  spectrum are obtained by distributing the atoms over the symmetric and antisymmetric single particle orbitals in the first band. This leads to $N+1$ energy levels, $N$ being the number of bosons.
$E_m = E_0 + m{\Delta}^0$ with $m =0,...,N$ where ${\Delta}^0 = \epsilon_1 - \epsilon_0$ is the energy difference between the two single particle orbitals in the first band. Thus for $g_0 =0$ the levels are equidistant (Fig.\ref{cap:spectrum}(a) inset) and we see  Rabi oscillation with frequency ${\omega_0}_1 = {\omega_1}_2 = {\Delta}^0$.
As the interaction is increased  ($g_0 = 0.005$), this equidistance is slightly broken (${\omega_0}_1 \simeq {\omega_1}_2$) and we get a superposition of two very close frequencies. This results in the formation of the beat pattern seen in the dynamics for $g_0 = 0.005$.

To understand the dynamics in the low interaction regime, it is instructive to map our system to a two-site Bose-Hubbard Hamiltonian \cite{Jaksch98,Fisher89}
\begin{equation}
 \hat{H} = -J(\hat{c}^{\dagger}_L\hat{c}_{R} + \hat{c}^{\dagger}_R\hat{c}_{L}) +  \sum_{j=L,R} \frac{U_j}{2}\hat{n}_j\left(\hat{n}_j-1\right)
\end{equation}
where  $J$ is the tunneling coupling, $U_{L,R}$ is the on-site energy of the left/right well and $\hat{n}_j\equiv\hat{c}_j^{\dagger}\hat{c}_j$.

 Before proceeding, we note here  that  there is no direct connection between the time-dependent SPF  used within our numerical M.C.T.D.H.  calculations and the parameters of the Bose-Hubbard (B-H) Hamiltonian. In the standard B-H model (which is valid in the weak interaction regime) the parameters $J$, $U_L$, $U_R$ are time independent constants while the shape of the orbitals such as the localized Wannier functions retain the shape throughout the course of the dynamics. Moreover, even for low energies the two most occupied modes for propagation do not necessarily coincide with the two modes of the B-H model. In this weak interaction regime, the B-H model is just a good approximation to our exact calculation and thus we have used it as solely an explanatory tool to analyze the results. 

Using the B-H Hamiltonian for $U_L,U_R\gg J$, the highest two eigenvalues are approximately $U_R$ and $U_L$. Whereas in the homogeneous case $\alpha = 0$ these two levels are close to  degenerate $U_L \approx U_R$ (Fig.\ref{cap:spectrum}(b)), here we have a breaking of the parity symmetry since $U_R >  U_L$ (Fig.\ref{cap:spectrum}(c)). This is understandable since two particles localized in the left well have lower energy than two particle in the right well leading to the energy level separation seen in Fig.\ref{cap:spectrum}(c). In terms of the number-state representation in the localized basis $\vert {N _L}^{(0)} ,{N_R}^{(0)}\rangle$ the degenerate eigenstates for the homogeneous case read

\begin{center}$\phi_{1,2}\approx \frac{1}{\sqrt{2}} (|0,2\rangle \pm |2,0\rangle $) \end{center}
and consequently the dynamics consists of shuffling the probability between the two states corresponding to a complete two particle tunneling.

In the case of sufficiently strong inhomogeneous interaction, the removal of the degeneracy of the energy levels leads to a decoupling of the  eigenstates into localized number-states
\begin{center}$\phi_1\approx  |2,0\rangle$ , $\phi_2\approx  |0,2\rangle $ \end{center}
This implies  that the initial state  $\psi(t=0) = \vert 2 ,0\rangle$ is very close to the 1st excited state $\phi_1$ and thus is effectively a stationary state of the system. This results in the suppression of tunneling for corresponding values of $g_0$

In the fermionization limit ($g_0\rightarrow \infty$) the system possesses the same local properties as a system of non-interacting fermions due to the Bose-Fermi mapping \cite{girardeau60}. Thus in an ideal case the inhomogeneity doesn't manifest ($g_{\pm}\rightarrow \infty$) and the tunneling dynamics is identical to a system of free fermions. As an idealization if we consider the initial state as two non-interacting fermions in the left well, then they would occupy the lowest two orbitals localized in the left well. In terms of the single particle eigenstates of the double well $| n_{a_{\beta}}^{(\beta)} \rangle$ where $n_{a_{\beta}}^{(\beta)}$ denotes the occupation number of the symmetric
($a_{\beta}=0$) or antisymmetric ($a_{\beta}=1$) orbital in band
$\beta$, the tunneling frequencies ${\omega}_{nn'} = E_n - E_n'$ are given by \cite{zoellner08}

\begin{equation} {\omega}_{nn'} = \sum_\beta \Delta^\beta\underbrace{({n_1}^\beta - {n'_1}^\beta)}_{= 0,\pm 1}\label{eq:fermi-frequ}\end{equation}
where $\Delta^\beta$ denotes the energy splitting of the band $\beta$ , ${n_1}^\beta$ represents the occupation of the anti-symmetric orbital of the band $\beta$.
Thus for two particles the contributing frequencies are  the lowest band Rabi frequency ${\Delta}^0$ and  the tunnel splitting of the first excited band ${\Delta}^1$. The tunneling dynamics can be pictured roughly as two fermions tunneling independently in the first two bands.

In our system however the finiteness of the $g_0$ value leads to deviations from the ideal fermionic dynamics. The inhomogeneity of the interaction still manifests leading to a difference w.r.t  the localized two-particle energy level in each well and the tunneling remains incomplete.

\subsection{Dynamics with varying inhomogeneity\label{sub:inhomogeneity}}

\begin{figure}
\begin{center}
\includegraphics[width=0.9\columnwidth,height=4.8cm]{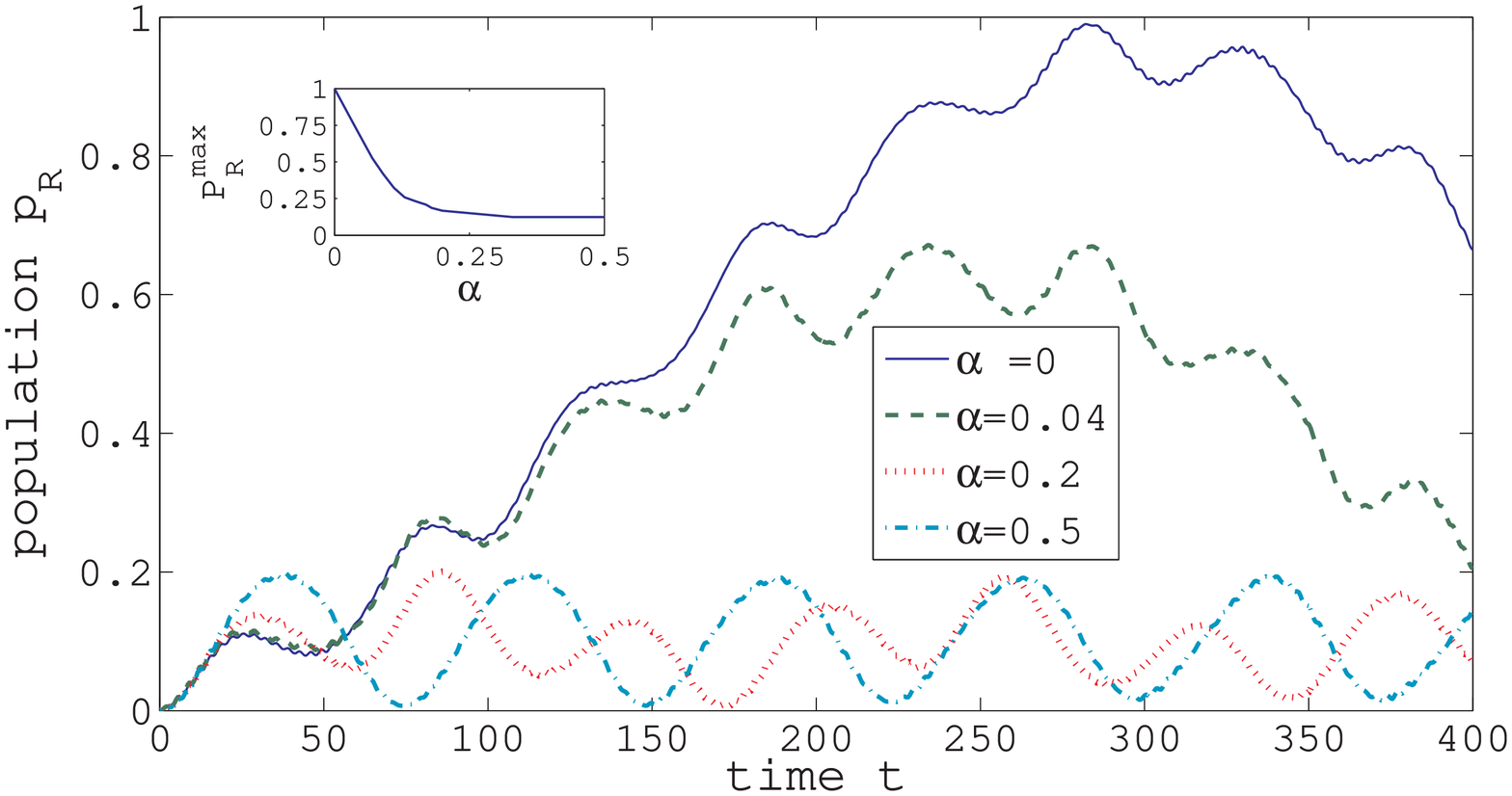}

\caption{(color online): Population of the right-well over time, $P_{\mathrm{R}}(t)$, at $g_0=0.2$ for different $\alpha$ values. \textit{Inset}: Variation of maximum population of the right well ${P_R}^{max}$ with $\alpha$ for $g_0=0.2$. \label{cap:2p_alpha_var} }
  \end{center}
\end{figure}

Having analyzed how the dynamics varies with changing interaction strength at a fixed interaction asymmetry, it is worthwhile to study the dependence of the tunneling dynamics  on the strength of the inhomogeneity. For this we study the effect of different $\alpha$ values on the tunneling dynamics for a fixed $g_0 = 0.2$.

In Fig. \ref{cap:2p_alpha_var} we observe that for $\alpha = 0$, we have complete tunneling with a two mode dynamics i.e.   fast oscillations (${\omega_0}_1$) which modulate slower tunneling oscillations (${\omega_1}_2$). When $\alpha$ is increased to a value of $0.04$, the tunneling maximum is reduced to roughly $0.7$ while still retaining the two-mode character. As $\alpha$ is further increased to $0.2$ the tunneling is suppressed as described in Sec. \ref{sub:analysis}. The characteristic display of fast and slow oscillations arising due to the time-scale difference of the contributing frequencies is not prominent here and for higher interaction asymmetry ($\alpha = 0.5$) we have effectively single mode tunneling with frequency ${\omega_0}_1$.

The variation of the maximum population ${P_R}^{max}$ with the inhomogeneity $\alpha$ (Fig.\ref{cap:2p_alpha_var} inset) shows a sharp drop with increasing $\alpha$ before effectively reaching a constant value $\sim 0.12$ for $\alpha \geq 0.3$. The reader should note that ${P_R}^{max}$ does not go to zero in the asymptotic limit $\alpha \rightarrow 1$ or $\frac{U_R}{U_L}\rightarrow\infty$. This is due to the fact that with a finite value of $g_0$ and a finite barrier height the tunneling coupling ($J$) is not negligible compared to $U_R$. As a consequence there remains a finite probability of bosonic tunneling between the two wells.

\subsection{Strong interaction inhomogeneity\label{sub:high}}

\begin{figure}
\begin{center}

\includegraphics[width=0.85\columnwidth,height=4.5cm]{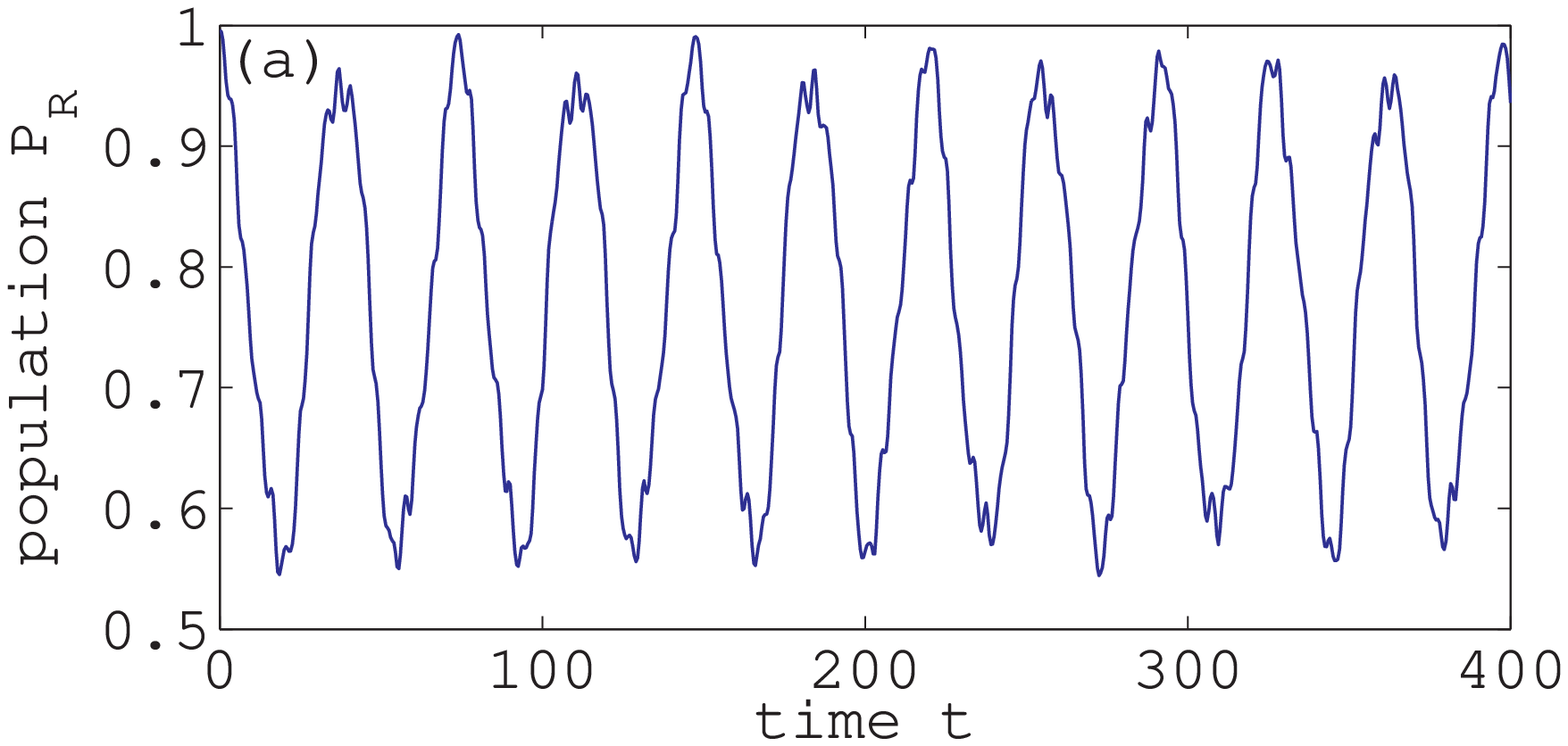}

\includegraphics[width=0.85\columnwidth,height=4.3cm]{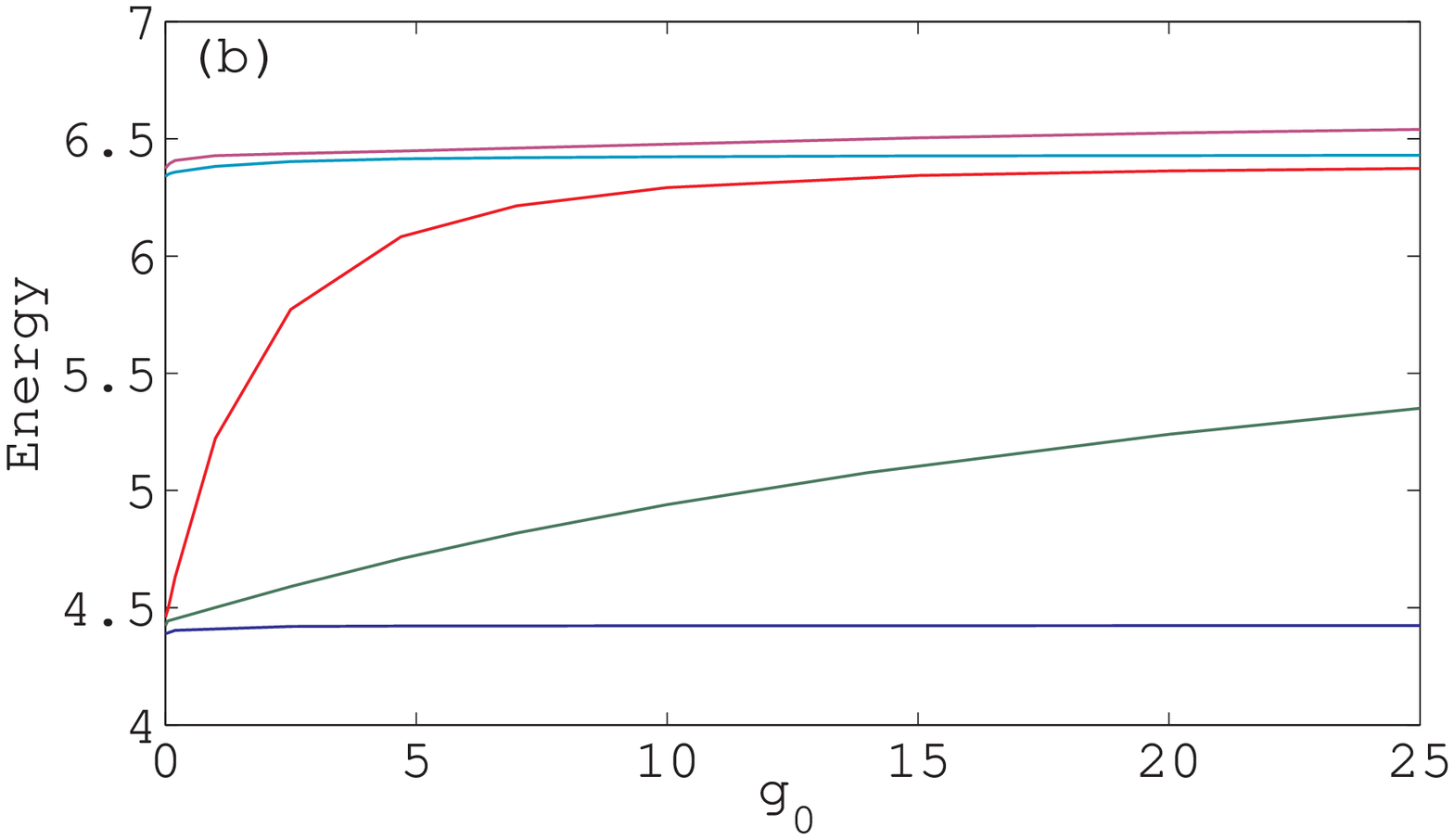}

\caption{(color online): (a) Population variation with time  $P_R(t)$ at $g_0 = 25$ and $\alpha = 1$ for $P_R(0)=1$, i.e initially populating the right-well. (b) Energy spectrum for $\alpha = 1$ \label{cap:alpha1} }
\end{center}
\end{figure}

An extremely strong inhomogeneity at a high interaction value leads to an  interesting higher band tunneling dynamics. We can realize such a system by having $\alpha = 1$ at $g_0 = 25$. This set up effectively makes the bosons fermionized in the right-well and almost non-interacting in the left. Preparing the initial set-up with both bosons  in the  left well leads to the  suppression of tunneling. However if we prepare the initial state with two boson in the right well, then we observe substantial tunneling. In  Fig.\ref{cap:alpha1} (a)  we see that the $P_R$ oscillates between 1 and 0.5 indicating a single boson tunneling with a single dominant frequency.  

In order to understand the phenomenon we look at the energy spectrum at $\alpha =1$ (Fig.\ref{cap:alpha1} (b)). While the ground-state remains unaffected, what we see is that close to the fermionization regime ($g_0 = 25$), the first excited state decouples from the higher three states which  come closer. The main contribution to the first excited state is the  state $|2,0\rangle$ and its separation from the other states could be understood from the fact that two boson in the left-well is almost non-interacting and thus energetically far off resonant from two effectively fermionized boson in the right-well $|0,2\rangle$. The consequences of this fact are the following: (i) The initial configuration of $|2,0\rangle$  becomes a stationary-state resulting in a highly suppressed tunneling, and (ii) the state $|0,2\rangle$ of the lowest band becomes energetically resonant and couples to the states $|1^1,1^0\rangle$ and $|1^0,1^1\rangle$  in the higher bands (where the superscript refer to the ground ($0$) or excited ($1$) orbital of the corresponding well). The latter leads to a tunneling dynamics in the higher band states predominantly between the 2nd and the 4th excited eigenstates (see Fig. \ref{cap:alpha1} (b))  which have greater overlap with the  initial state $|0,2\rangle$. These orbitals have mostly contributions from the states $|0,2\rangle$ and $|1^1,1^0\rangle$ while the other orbital has minimal overlap with the initial state. As a result we get a single-particle tunneling with one dominant frequency given by the splitting of the energy between these two levels. In other words, we effectively have a single boson tunneling between the wells in the excited band. Note that this highly correlated single-particle tunneling scenario is  attributed to the high inhomogeneity in the strong interaction regime since the combination of these two factors are responsible for turning the pair-tunneling scenario off-resonance.

 \section{Multi-Particle Dynamics \label{sec: many particle}}

\begin{figure}

\includegraphics[width=0.85\columnwidth,keepaspectratio]{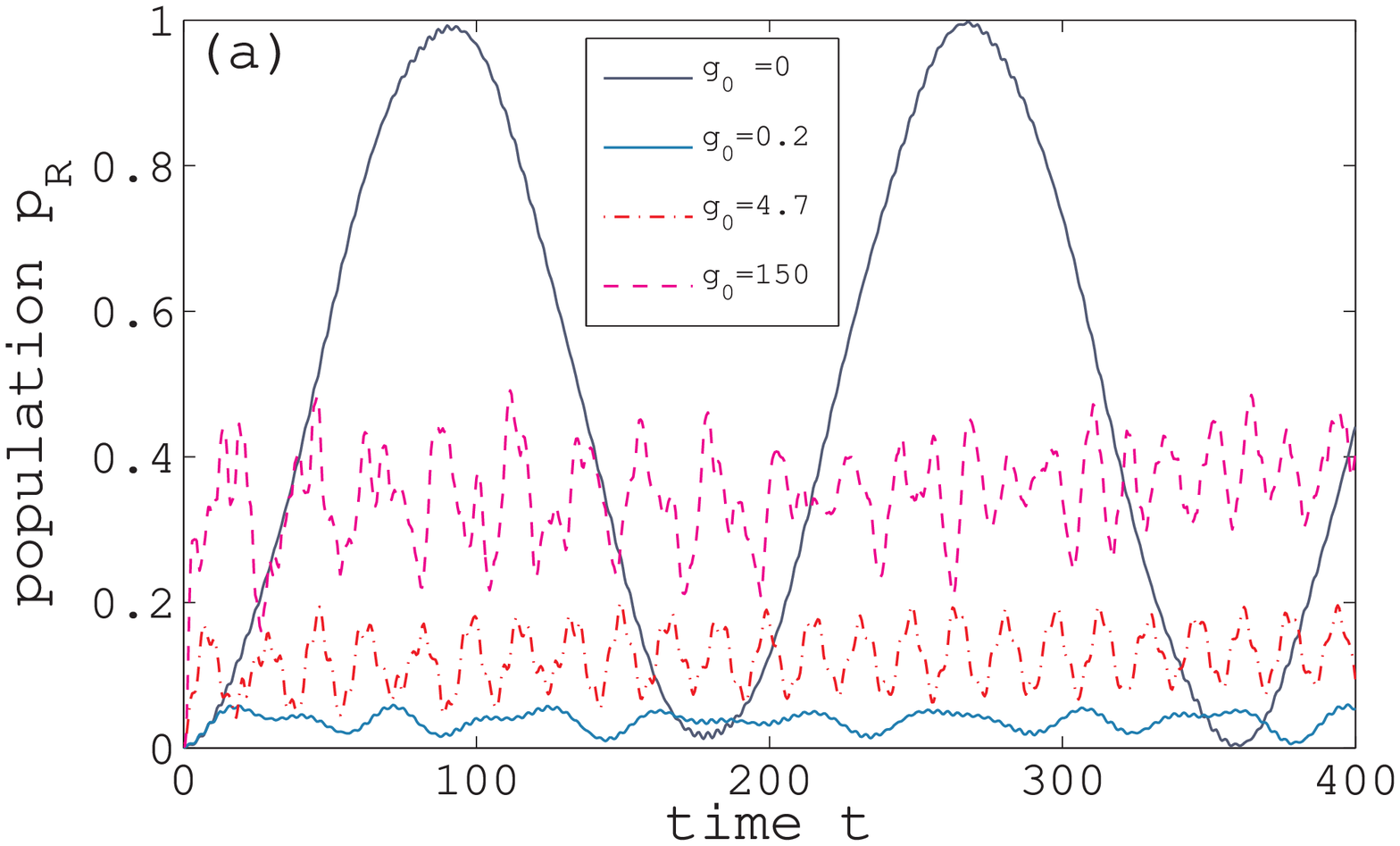}

\includegraphics[width=0.85\columnwidth,keepaspectratio]{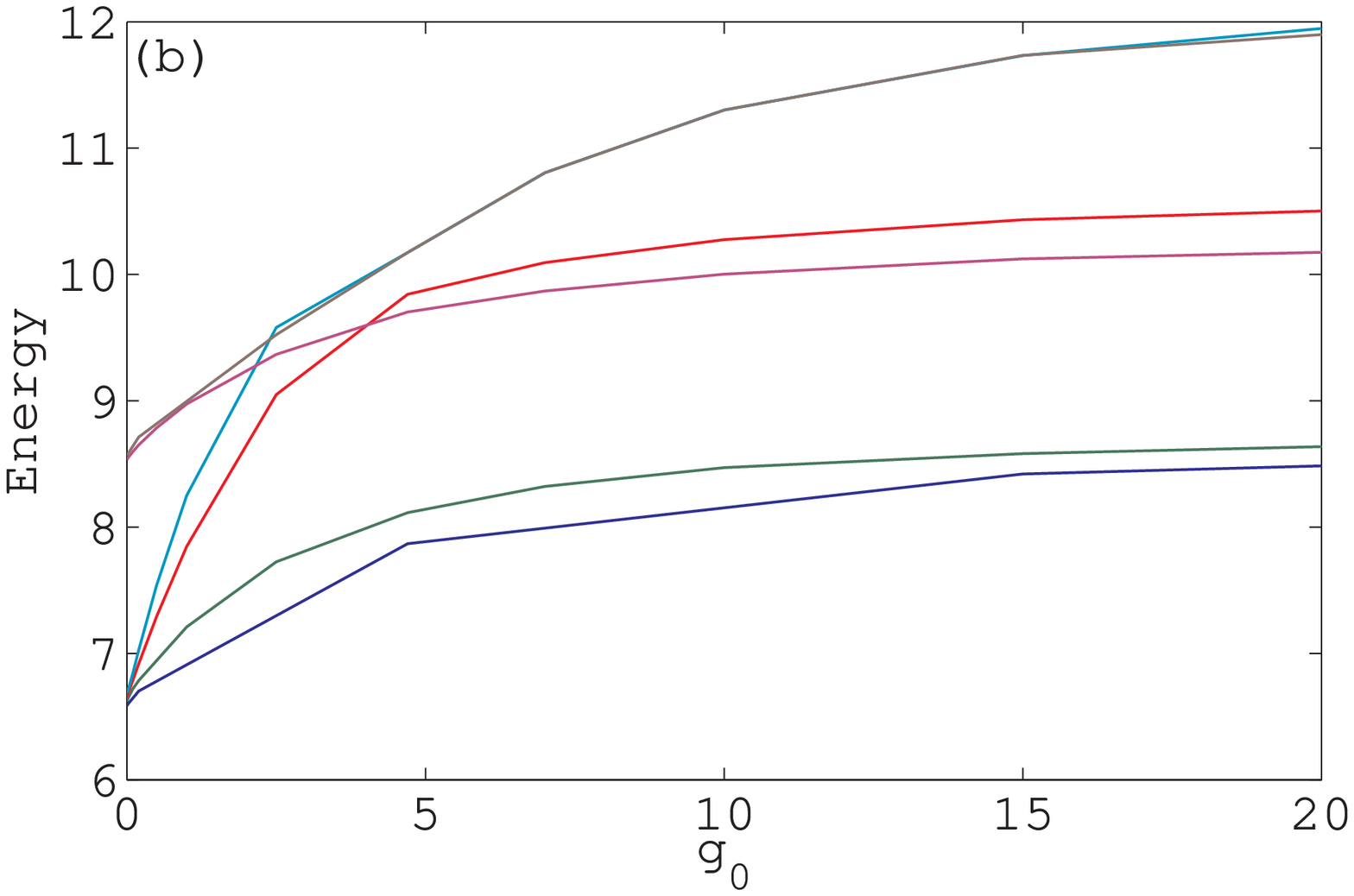}

\caption{(color online)(a) Population of the right-hand
well over time, $P_{\mathrm{R}}(t)$, for three bosons for different interaction strengths at $\alpha = 0.2$. (b) Three boson energy spectrum at $\alpha = 0.2$. \label{cap:3p_nr}}
\end{figure}

 Having analyzed the tunneling dynamics of two atoms let us now focus on the case of three or more atoms to see the general atom number dependence of tunneling in the presence of spatially modulated interactions.

\subsection{General behavior and mechanisms \label{sub:3p general}}

Like in the two boson case we start with the initial state of $N = 3$ bosons prepared in the left well. As shown in Fig. \ref{cap:3p_nr}(a), the main effects are similar to the two-atom case. The dynamics is again governed by frequencies determined by the energy difference of the low lying spectrum.  For very small interaction, the nearly equal energy difference gives rise to the beat pattern similar to that of two particles. As we increase the  interaction strength, we observe suppression of tunneling for $g_0 = 0.2$ followed by a partial restoration at $g_0 = 4.7$ and a higher amplitude reemergence close to the fermionization limit at $g_0 =150$.
The general mechanism for the suppression is the same as for the two particle case. Now, however, in the symmetric case $\alpha = 0$, the contributing nearly degenerate eigenstates are of the form  $|N,0\rangle \pm |0,N\rangle$. Consequently we have a complete $N$ particle tunneling with a frequency given by \cite{salgueiro06} $\omega\sim2NU/(N-1)!\times(2\Delta^{0}/U)^{N}$ where $U = U_L,U_R$ denotes the on-site interaction energy. The tunnel period thus grows exponentially with $N$.
When the inhomogeneous interaction is introduced, the states decouple to the localized number-states $|N,0\rangle$ and $|0,N\rangle$ and thus the initial state becomes a stationary one leading to the suppression of tunneling.
The important thing to note is that with increasing $N$, the suppression of tunneling occurs for much smaller values of $g_0$. For instance at $g_0 =0.2$ for $N=3$ we have almost complete suppression in contrast with $N=2$ where we still observed  significant tunneling (see Fig.\ref{cap:2p_g_var}) for this value of $g_0$. This could be understood from the fact that the contribution of the on-site energy on the cat-state goes as $\sim U_{L,R}N(N-1)/2$, while that of the tunneling term is $N$ independent. This fact is responsible for a significant decoupling of these states at a lower $g_0$ value leading to faster suppression of tunneling as $N$ increases.

Also unlike that of the two boson case, the spectrum for the three boson case contains crossings between the higher-lying states (see Fig.\ref{cap:3p_nr}(b)) and in the vicinity of these crossings there is a partial reappearance of tunneling. This can be seen for instance at $g_0=4.7$ where we observe a restoration in the three-particle case whereas for two particles we still observed a significant suppression (see Fig. \ref{cap:2p_g_var}). In this regime the higher bands contribute more significantly leading to the convoluted dynamics observed. These higher band contributions leads to further recovery with increasing interaction strength towards the fermionization regime although even for $g_0 = 150$  we do not get the exact fermionic dynamics which is characterized by the tunneling of three independent fermions.

\subsection{Generating tunneling resonances by interaction inhomogeneity
\label{sub: asymmetry tunneling resonance}}

\begin{figure}

\includegraphics[width=0.85\columnwidth,height=4.5cm]{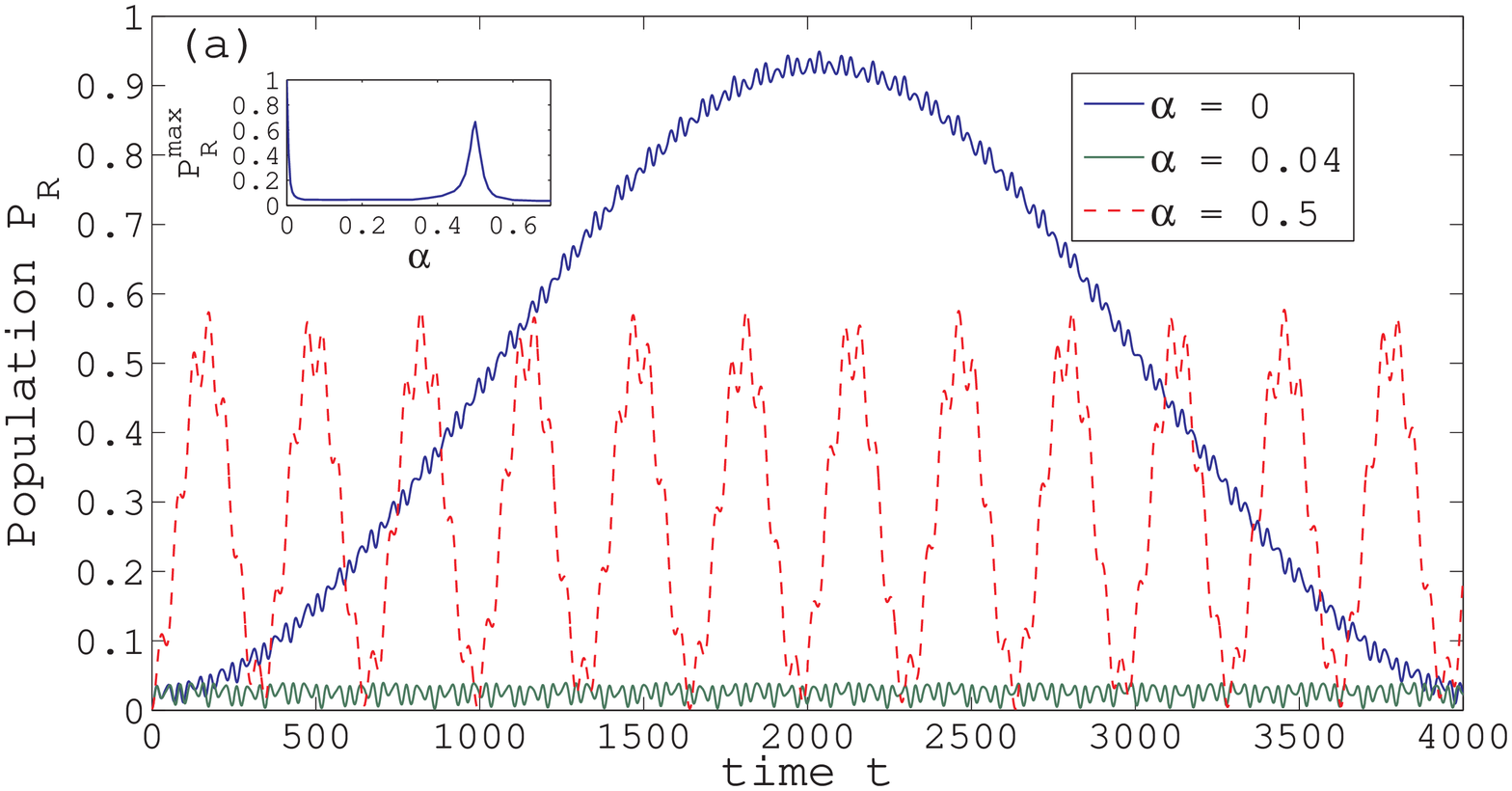}

\includegraphics[width=0.85\columnwidth,height=4.5cm]{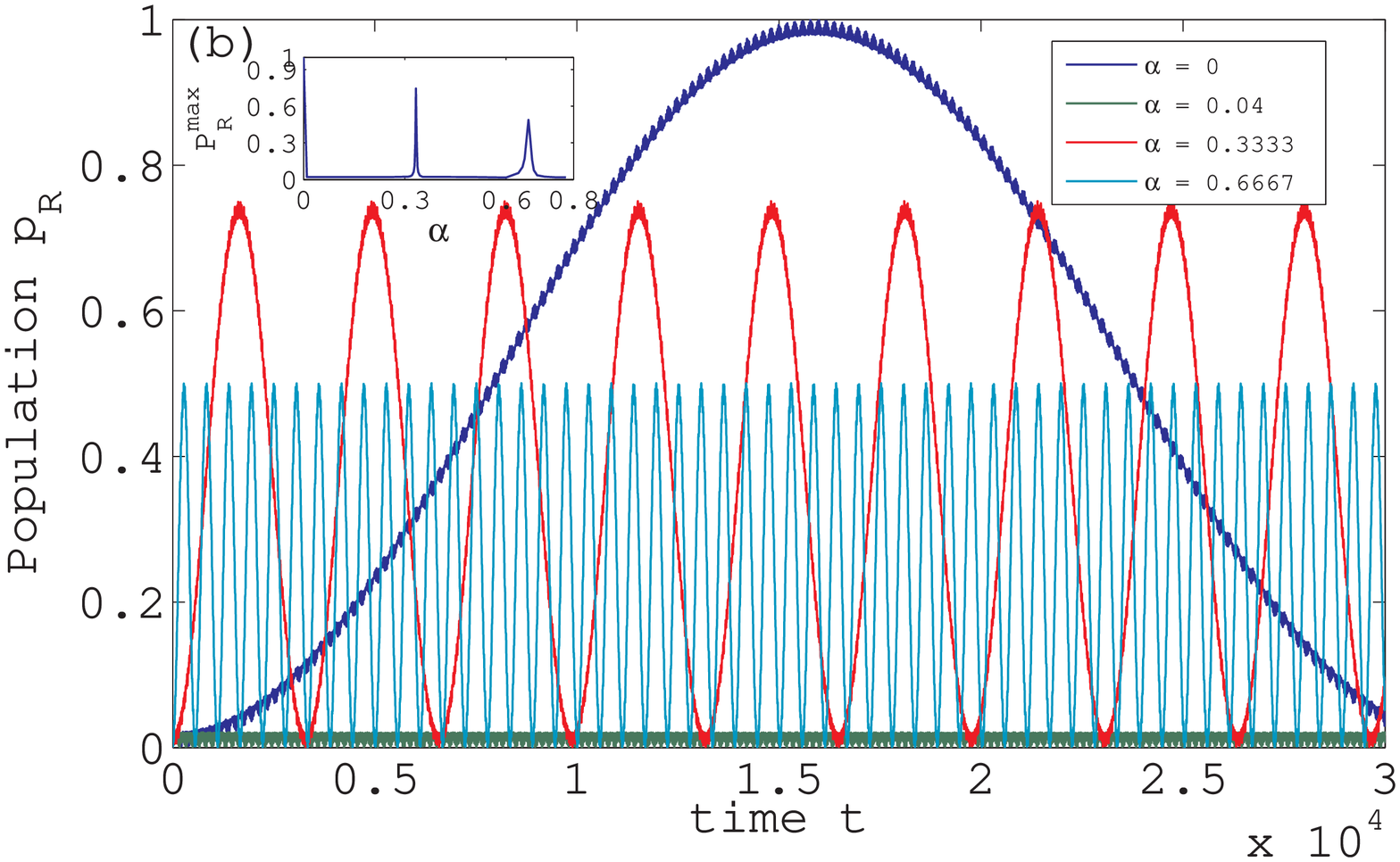}

\caption{(color online): Population of the right-well over time, $P_{\mathrm{R}}(t)$, at $g_0=0.2$ for different $\alpha$ values for (a) 3-particles and (b) 4-particles. \textit{Inset}: Variation of maximum population of the right well ${P_R}^{max}$ with $\alpha$ for $g_0=0.2$. \label{cap:3p_pr_alpha} }
\end{figure}

A very interesting phenomenon for the $N \geq 3$ particle case  is that by tuning the asymmetry $\alpha$ we get a controllable reemergence of tunneling. To observe this, we study how the tunneling dynamics changes  with different values of $\alpha$ for  $g_0 = 0.2$ (Fig.\ref{cap:3p_pr_alpha}). The value of $g_0$  is chosen such that  the inhomogeneity effect manifest but is still in the two-mode regime. For three atoms we observe (Fig.\ref{cap:3p_pr_alpha}(a)) that a complete tunneling for $\alpha = 0$ gives way to suppressed tunneling with increasing $\alpha$ value. However at $\alpha = 0.5$ we observe a reappearance which is in form of a tunneling resonance peaked at $\alpha = 0.5$ with ${P_R}^{max} \approx 0.6$ corresponding to effective two boson tunneling.
In the case of $N=4$ we see two resonances (fig.\ref{cap:3p_pr_alpha}(b)\textit{inset}) - the larger one centered on $\alpha = 0.3333$ with an amplitude $0.75$ and the smaller one at $\alpha = 0.6667$ with an amplitude  $0.5$ resulting in the reappearance of tunneling shown in  Fig.\ref{cap:3p_pr_alpha}(b).

\begin{figure}

\includegraphics[width=6.5cm, height=3.5cm]{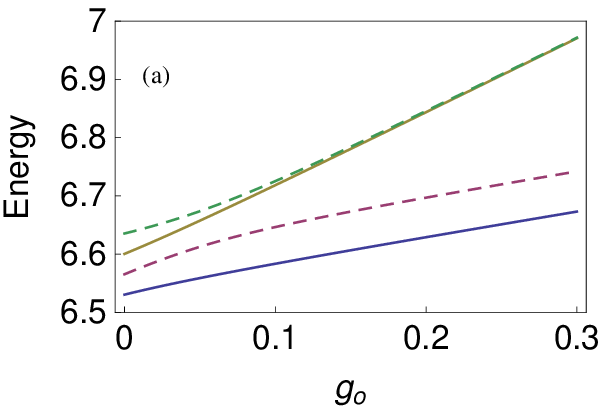}

\includegraphics[width=0.75\columnwidth,height=3.5cm]{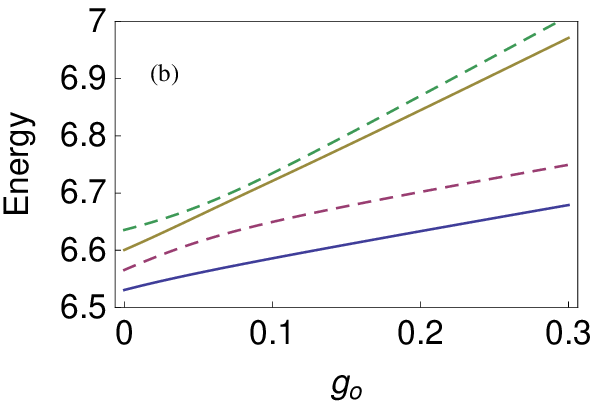}

\includegraphics[width=0.75\columnwidth,height=3.5cm]{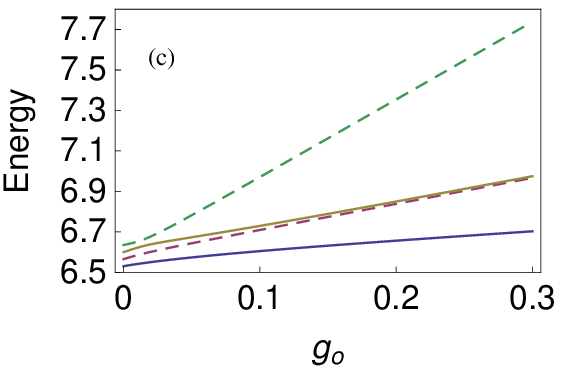}

\caption{(color online):Three particle energy levels for $0<g_0<0.3 $ for (a) $\alpha = 0$, (b) $\alpha = 0.04$ and (c) $\alpha = 0.5$.
\label{cap:3p_spec}}

\end{figure}

In order to understand this we have to study the spectra and the underlying eigenstates for different $\alpha$ (Fig.\ref{cap:3p_spec}). In the case of $N=3$ for no  asymmetry $\alpha = 0$, the highest two levels form a doublet (Fig.\ref{cap:3p_spec}(a)) and the corresponding eigenstates are degenerate of the form $\frac{1}{\sqrt{2}}(|3,0\rangle \pm |0,3\rangle)$. As $\alpha$ is increased the parity symmetry is broken and the doublets separate and likewise the eigenstates decouple (Fig.\ref{cap:3p_spec}(b)). The energy eigenvalues (in the limit of very high $g_0$) are given by $U_L$, $U_R$, $3U_L$ and $3U_R$ with the corresponding eigenstates  $|2,1\rangle$, $|1,2\rangle$, $|3,0\rangle$ and $|0,3\rangle$. However, when $U_R \approx 3U_L$ ($\alpha = 0.5$) the 1st and the 2nd excited eigenstates become near degenerate and form a doublet of the form $\frac{1}{\sqrt{2}}(|1,2\rangle \pm |3,0\rangle)$ (Fig.\ref{cap:3p_spec}(c)). Thus the initial state $|3,0\rangle$ is no longer a stationary state of the  system. As a consequence we get a restoration of tunneling and the dynamics basically involves shuffling atoms between these two number-states. In other words we have tunneling of two particles  between the two wells while one particle remains in the left well. This resonant two particle tunneling is what we observe for the $\alpha = 0.5$ case. As $\alpha$ is increased further this degeneracy is once again broken and the states decouple leading back to the suppressed tunneling dynamics. This is reminiscent of what happens in the asymmetric double-well for homogeneous interactions \cite{dounasfrazer07a}.

In similar consideration, for the 4-particle case the energy eigenvalues are $3U_L$, $6U_L$, $(U_L + U_R)$, $3U_R$ and $6U_R$. Now if $U_R \rightarrow 2U_L$ ($\alpha = 0.3333$) then we have two degeneracies viz  $3U_R \rightarrow 6U_L$ and $(U_L + U_R) \rightarrow 3U_L$ corresponding to the eigenstates $\frac{1}{\sqrt{2}}(|4,0\rangle \pm |1,3\rangle)$ and $\frac{1}{\sqrt{2}}(|3,1\rangle \pm |2,2\rangle)$. Since the initial state is $|4,0\rangle$ only the first degeneracy contributes. Thus the dynamics in this case consists of  tunneling of three bosons between the wells while one boson remains in the left well. This results in the tunneling amplitude of $0.75$.
The second tunneling peak occurs for $U_R \rightarrow 5U_L$ ($\alpha = 0.6667$) which leads to $(U_L + U_R) \rightarrow 6U_L$. The corresponding degenerate eigenstates are $\frac{1}{\sqrt{2}}(|4,0\rangle \pm |2,2\rangle)$ and we observe  tunneling of two bosons on top of others remaining in the left well  and thus the tunneling peak of $0.5$.
The above analysis can be extended  generically for $N$ particles where we would have $N-2$ resonances corresponding to the degeneracies between the eigenstates.

\subsection{Correlations\label{sub:correlations}}

\begin{figure}

\includegraphics[width=0.75\columnwidth,keepaspectratio]{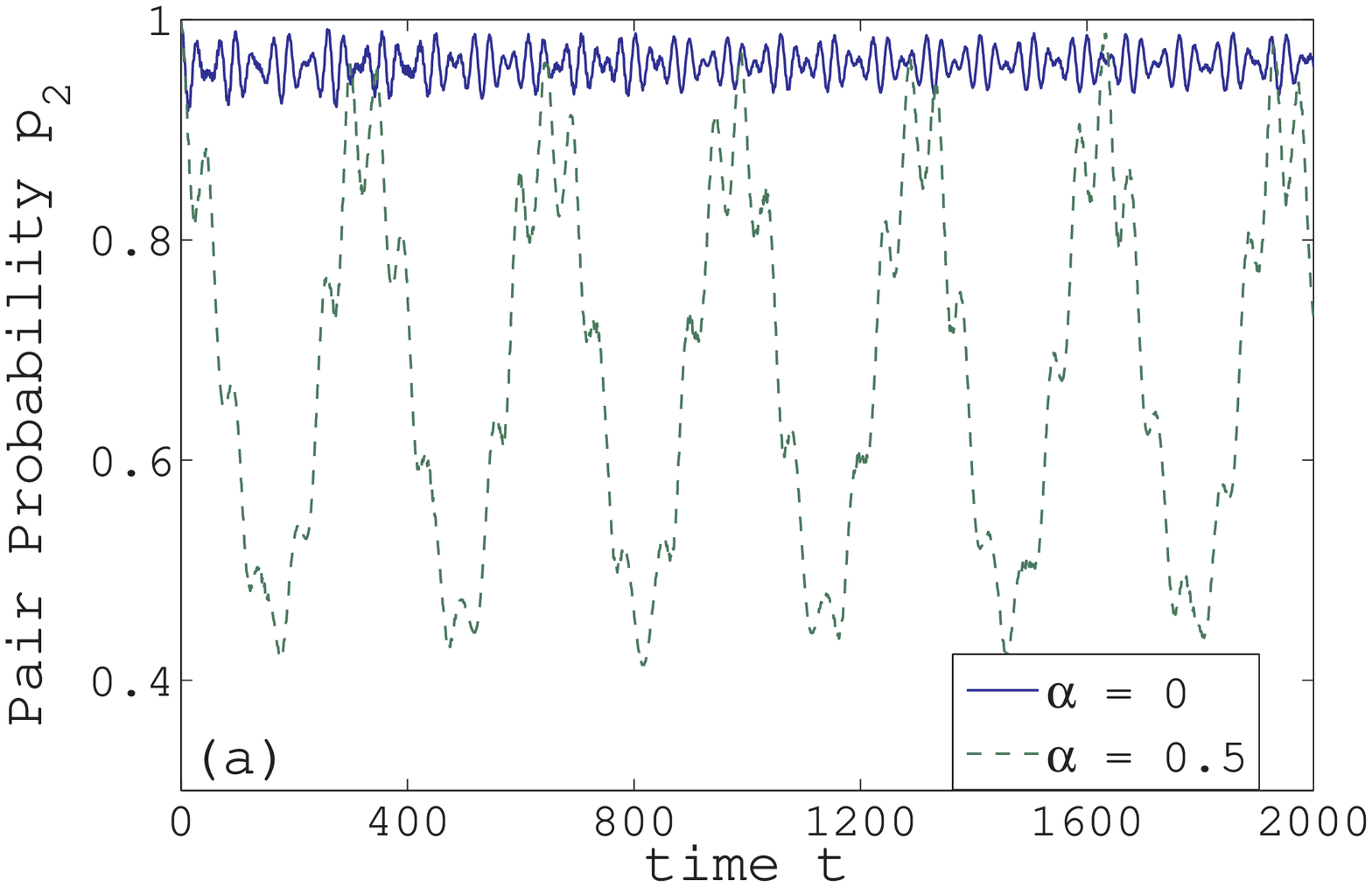}

\includegraphics[width=0.75\columnwidth,keepaspectratio]{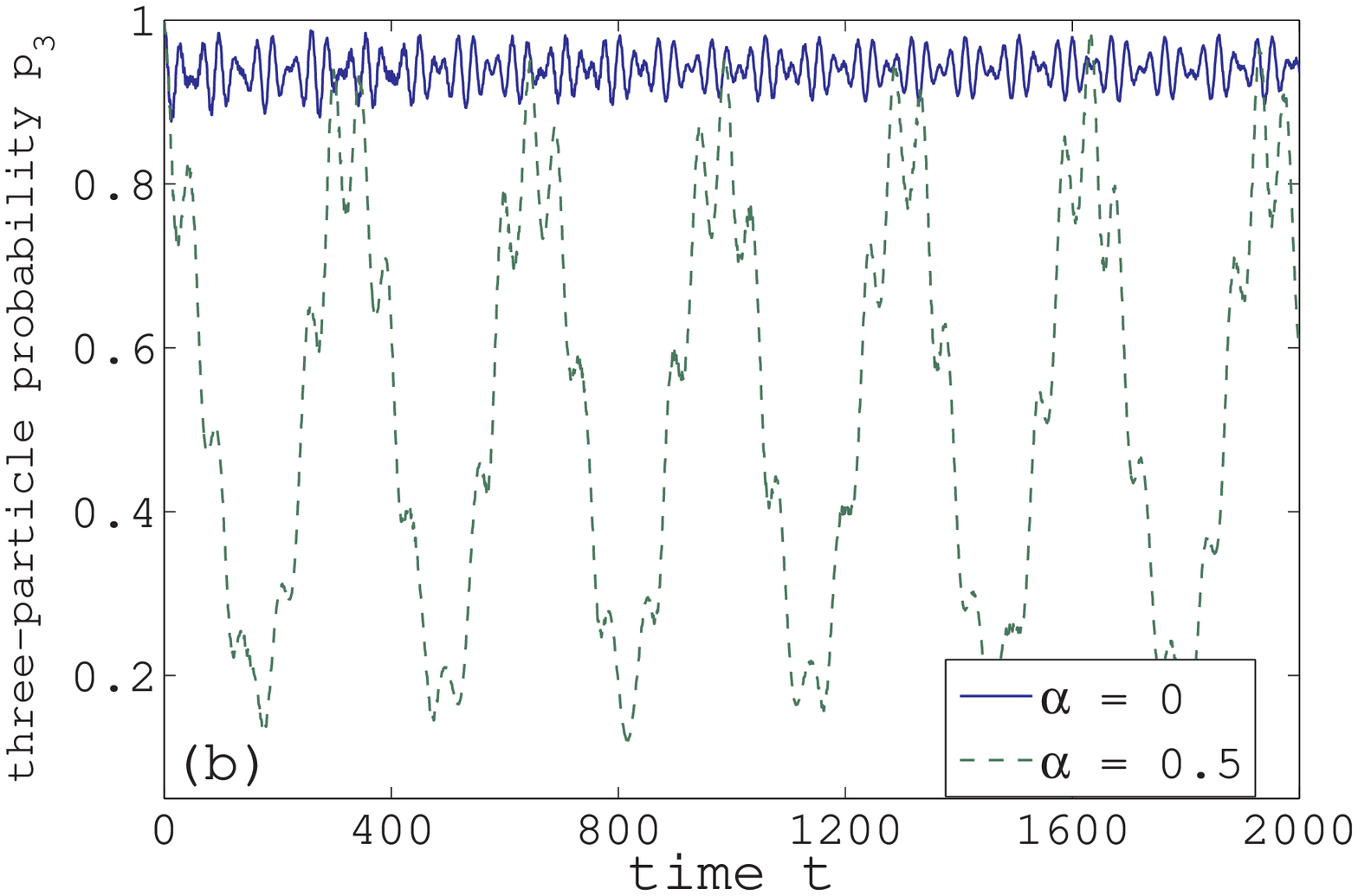}

\caption{(color online) Temporal evolution of  (a) pair-Probability   and (b) three particle probability  at $\alpha =0$ and $\alpha = 0.5$ for $N = 3$ and $g_0 = 0.2$ . \label{cap:3p_correlation}}
\end{figure}

In order to study the exact nature of tunneling dynamics, we need to investigate the correlations between the particles. For this we study the temporal evolution of the pair-probability or the probability of finding two particles in the same well defined by

\begin{equation}
p_{2}(t)  =  \langle\Theta(x_{1})\Theta(x_{2})+\Theta(-x_{1})\Theta(-x_{2})\rangle_{t}\\
\end{equation}
and the three-particle-probability or the probability of finding all three particles in the same well defined by

 \begin{equation}
p_{3}(t)  =  \langle\Theta(x_{1})\Theta(x_{2})\Theta(x_{3})+\Theta(-x_{1})\Theta(-x_{2})\Theta(-x_{3})\rangle_{t}\\
\end{equation}
In the case of $N=3$, for homogeneous interaction $\alpha = 0$ at $g_0 = 0.2$ both $p_2$ and $p_3$ oscillate close to unity (Fig.\ref{cap:3p_correlation}). This implies that all the three particles can be found in the same well or in other words they tunnel together between the wells. This confirms the analysis of the dynamics by the eigenstate analysis in the preceding section as tunneling between $|3,0\rangle$ and $|0,3\rangle$ states.

Similarly at resonance ($\alpha = 0.5$) we find that $p_3$ oscillates from 0.1 and 1 implying that the system oscillates between a three-particle state to a non-three-particle state, namely the pair-state $|1,2\rangle$ which can be inferred from the variation of $p_2$ (Fig.\ref{cap:3p_correlation}(b)). As a result we have pair tunneling on top of a particle remaining in the left-well. 
(Ideally in the case of B-H model, $p_2$ should be oscillating between 1 and 0.33 while $p_3$ between 1 and 0. However in our case the realistic potential and parameter regimes as well as some higher band contributions leads to the some deviations from this behavior).

\section{Asymmetric Double-Well\label{sec:asymmetric}}

Thus far we have investigated the dynamics in symmetric double-well with inhomogeneously interacting bosons. An interesting extension is to study the dynamics in an asymmetric double-well. This gives us the chance to examine the interplay between the interaction inhomogeneity and the tilt. A special interesting consideration would be to see if the tilt could be tuned to offset the inhomogeneity in the interaction and mimic the dynamics of symmetric interaction case or further if it can generate some new tunneling resonances.

\subsection{Generating tunneling resonances by a tilt.}

\begin{figure}

\includegraphics[width=0.75\columnwidth,height=4.3cm]{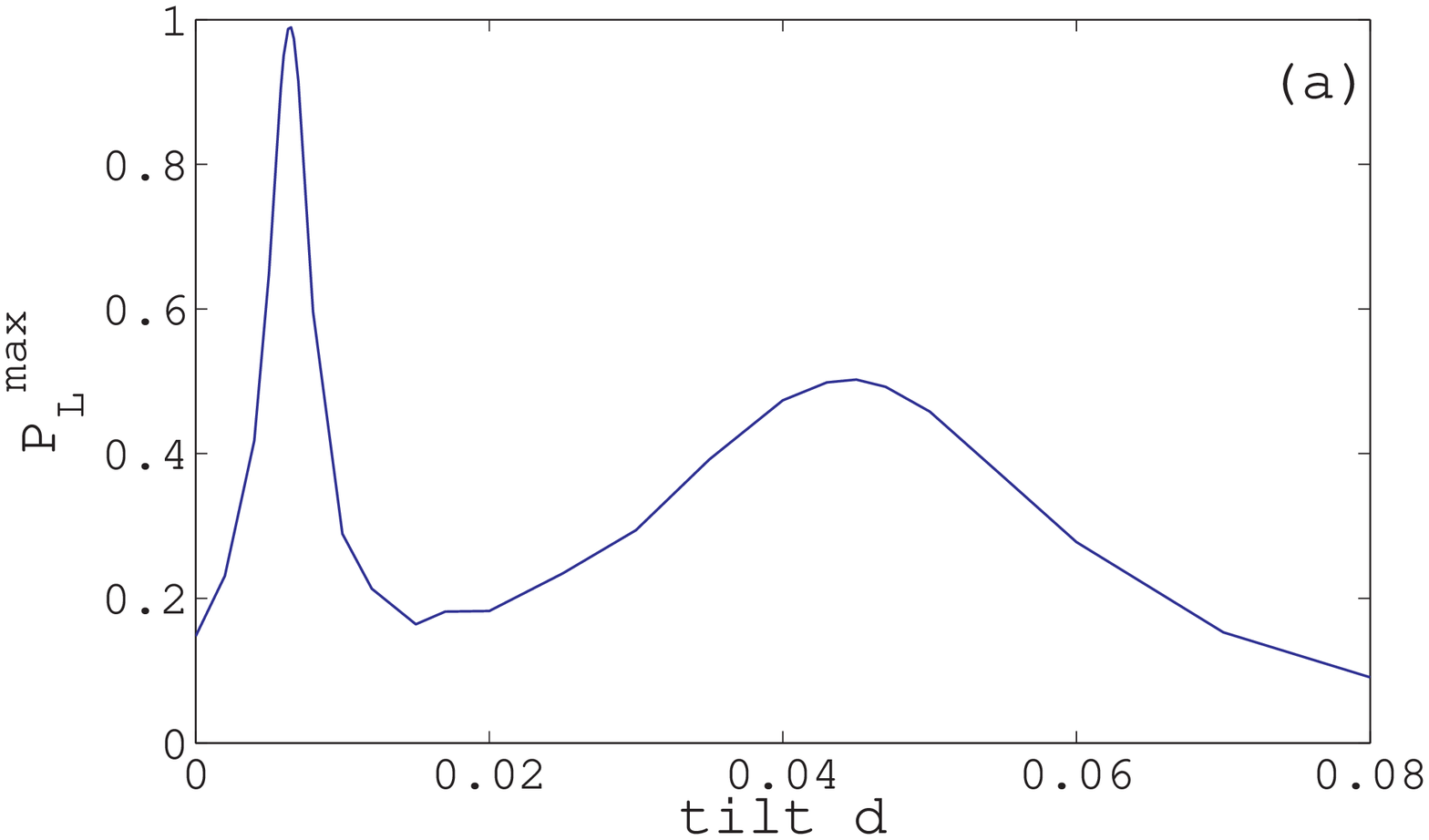}

\includegraphics[width=0.75\columnwidth,height=4.3cm]{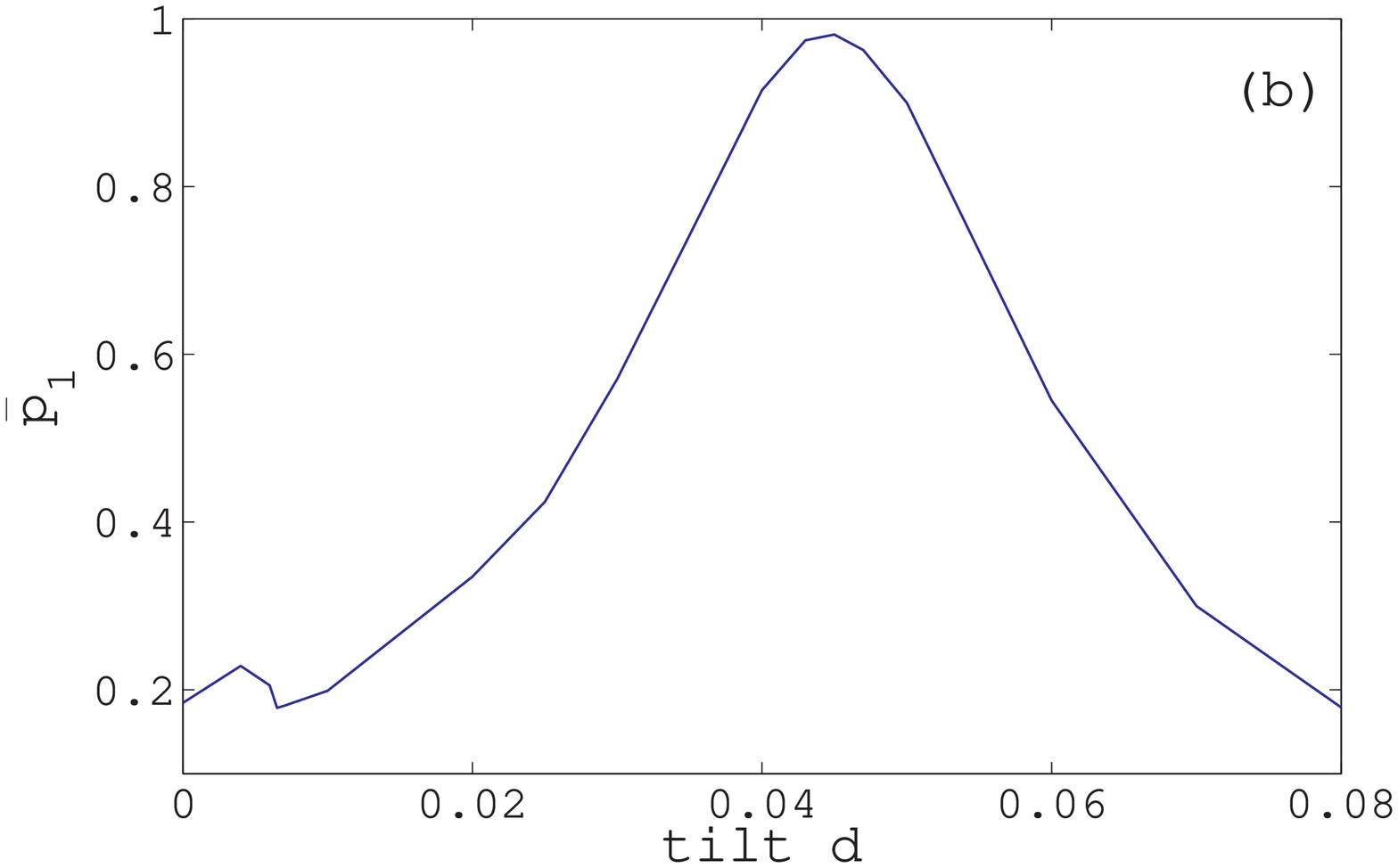}

\caption{Variation of (a) tunneling maximum ${P_L}^{max}$ with tilt $d$ (b) maximum single particle probability  $\bar{p}_{1}$ with tilt $d$ for $N = 2$, $g_0 = 0.2$ and $\alpha = 0.2$. \label{cap:2p_tilt} }
\end{figure}

In symmetric wells with homogeneous interaction, the localized $N$ particle state $|N,0\rangle$ has the same energy as that of the state $|0,N\rangle$  resulting in a complete $N$-particle tunneling between the wells. With the introduction of the inhomogeneity w.r.t the  interaction, this resonance is broken and the energy of $N$ particles in the right well is higher than that in the left well resulting in the suppression of tunneling as seen before. Now if we incorporate a tilt in the double well such that the left well is lifted and right well is pushed down energetically in exactly the right amount to make the localized $N$ particle energy levels resonant then we should expect a reemergence  of tunneling.

To observe this we  prepare the initial state with both particles in the right well $\psi(0) = |0,2\rangle$ and study the variation of the tunneling maximum  ${P_L}^{max}$ with a  tilt $d$ (Fig.\ref{cap:2p_tilt}(a)) incorporated into the Hamiltonian as a linear term $-dx$. We restrict ourselves to the $\alpha = 0.2$ and $g_0 = 0.2$ case. We observe a sharp resonance at $d \approx 0.0065$ corresponding to the tilt which exactly balances the localized pair-state energy difference due to inhomogeneous interaction. The result is pair-tunneling between the two wells as we would have it in a completely symmetric set-up.

With higher tilt the tunneling maximum falls off very sharply as the pair-state becomes off-resonant again and we get a suppression of tunneling. The next maximum occurs when the tilt is large enough to make the localized pair state $|0,2\rangle$ resonant with the state  $|1,1\rangle$. This results in a broad  tunneling maximum at $d \approx 0.045$ corresponding to single-particle tunneling.

 To confirm our analysis of the tunneling mechanism we look at the variation of maximum single particle probability $\bar{p}_{1}$ with tilt (Fig.\ref{cap:2p_tilt}(b)),  defined as $\bar{p}_{1}={\max}_t({1-p_{2}(t)})$  which gives the probability of having only one particle in a well. We observe a negligible value at the first resonance $d \approx 0.0065$ confirming that the dynamics is pair-tunneling  while a very broad maximum peaked at the second resonance $d \approx 0.045$ corresponds to the maximum probability of finding a single particle  which in our case is  the $|1,1\rangle$ state and the dynamics is a single particle tunneling between the $|0,2\rangle$ and $|1,1\rangle$ states.

\subsection{Spectral Analysis}

\begin{figure}
\includegraphics[width=0.75\columnwidth,keepaspectratio]{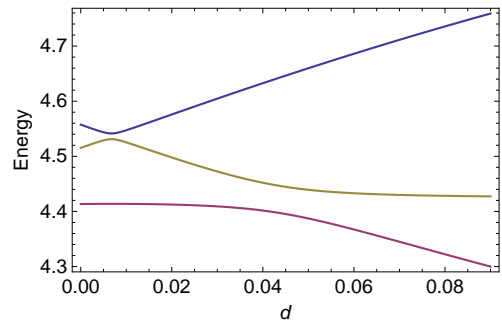}

\caption{(color online) Two particle energy spectrum with tilt $d$ for $\alpha = 0.2$ and $g_0=0.2$. \label{cap:tilt_spectrum}}
\end{figure}

To understand the effect of the tilt on the tunneling dynamics we study the energy spectra $E$ with varying tilt $d$ at fixed $g_0 =0.2$ and $\alpha =0.2$ (Fig.\ref{cap:tilt_spectrum}).
At $d = 0$ the eigenstates are basically number-states in the localized basis.  With increasing $d$, the highest two levels $|0,2\rangle$ and $|2,0\rangle$  move closer and form a sharp avoided crossing at $d \approx 0.0065$ corresponding to the first tunneling resonance. At this point the tilt exactly balances the interaction inhomogeneity and the eigenstate is in form of the cat-state $|2,0\rangle \pm |0,2\rangle$.  This state is very sensitive to the tilt and a minute perturbation decouples them into the localized number-state resulting in a very sharp tunneling resonance. The ground-state, which is the $|1,1\rangle$ state is insensitive to the tilt since this lowering of one particle and raising another particle keeps the state energetically unaffected within the linear regime. This state forms a broad (anti)crossing with the lower excitedstate at $d \approx 0.045$ forming the broad single-particle tunneling resonance seen in the dynamics. This behavior seen in the two-particle case can be expected in general for $N$  particles giving $N$ resonances corresponding to the avoided crossings encountered. In particular with increasing tilt, the successive resonances corresponds to a mechanism where one less particle tunnels compared to that of the previous one while the width of the resonances becomes progressively broader. 

\section{Conclusion and Outlook}

We have investigated the double-well tunneling dynamics with inhomogeneous interaction. More specifically we modeled the system such that we have two different interaction strengths in the two wells. What we observe is that this inhomogeneity leads to a suppression of tunneling. The reason for this suppression can be attributed to the breaking up of the doublet structure in the energy spectrum leading to a decoupling of the eigenstates into the localized number-state. Increasing the interaction to the fermionization limit leads to a reappearance of the tunneling. The dynamics is governed by the band splitting of the first two bands although the finiteness of the interaction strength and the presence of the interaction inhomogeneity leads to deviation from the ideal fermionic behavior. For a very pronounced  interaction inhomogeneity for strong  interactions, we observe  single particle tunneling between the localized excited bands of the double-well. 

These basic considerations can be used to understand the many particle system. There we observed a more severe suppression of tunneling for even lower $g_0$ values. Most importantly for $N\geq3$ atoms, one can generate tunneling resonances by tuning the interaction asymmetry. These resonances occur as a result of the  formation of degeneracies between different eigenstates. For three particles, the exact tunneling mechanism  was investigated using the evolution of the pair-probability and the three-particle probability. These studies show that we get correlated pair and triplet tunneling with a complete absence of single particle tunneling.

Finally we explored the dynamics in a asymmetric double-well and this gives us an understanding of the interplay between the interaction inhomogeneity and the tilt. We observe that the tilt can be tuned to offset the interaction inhomogeneity leading to a tunneling resonance. These dynamics have been explained through the spectral analysis in terms of avoided crossings between the levels. 
Note that an interesting prospective would be to try to describe the presently found effects in the context of a generalized Bose-Hubbard model, where the on-site energies and the coupling constants would be site- and occupation number dependent\cite{luhmann10, Schneider09, kaspar10}.

 Understanding the few-body mechanisms of tunneling with spatially modulated interactions can be used to design schemes for selective transport of particles between different wells and/or reservoir systems \cite{Schlagheck09,Schlagheck10}. Further our study could serve as a starting point for the investigation of the quantum dynamics in the presence of time-dependent interaction modulations and even be extended to multi-well systems \cite{alexej09}.

 \acknowledgments{ B.C gratefully acknowledges the financial and academic support from the  International Max-Planck Research School for Quantum Dynamics. Financial support from the  German Academy of Science Leopoldina (grant LPDS 2009-11) is gratefully acknowledged by S.Z. P.S. acknowledges financial support by the Deutsche Forschungsgemeinschaft.
The authors appreciate fruitful discussions with H.D. Meyers. }

\end{document}